\documentclass[preprint,showpacs,amsmath,amssymb,byrevtex,endfloats*,citeautoscript,altaffilletter,a4paper]{revtex4}

\usepackage{graphicx}
\usepackage{amssymb}
\usepackage{amsmath}

\begin{document}

\author{Angelo Facchini}
\email{a.facchini@unisi.it}
 \affiliation{Department of Chemical and Biosystems Sciences,
University of Siena\\
  Via della Diana 2/A 53100 Siena, Italy.}%Lines break automatically or can be forced wit

\author{Carlo V. Bellieni}
 \affiliation{University of Siena, Dept. of Pediatrics, Obstetrics and Reproductive Medicine,
                  Italy}

\author{Nadia Marchettini}
 \affiliation{Department of Chemical and Biosystems Sciences,
University of Siena\\
  Via della Diana 2/A 53100 Siena, Italy.}

\author{Federico M. Pulselli}
 \affiliation{Department of Chemical and Biosystems Sciences,
University of Siena\\
  Via della Diana 2/A 53100 Siena, Italy.}

\author{Elisa B.P. Tiezzi}
 \affiliation{University of Siena, Dept. of Mathematics and Informatics, Italy}

\title{Relating pain intensity of newborns to onset of nonlinear phenomena in cry recordings}

\begin{abstract}
The cries of several full term newborns, recorded during blood
sampling, were analyzed. Spectrograms showed the appearance of
irregular patterns related to the pain assessed using the method of
the DAN scale \cite{carbajal97}. In particular, the appearance of
Noise concentration Patterns (NP) in spectrograms was related to the
increase of the pain suffered by the newborns. In this scenario,
pain constitutes a bifurcation parameter for the vocal folds
dynamic, inducing a Ruelle-Takens-Newhouse chaotic transition.
\end{abstract}


 \maketitle

% main text
\section{Introduction}
The vocal folds, together with glottal airflow, constitute a highly
nonlinear oscillator. The vocal folds are set into vibration by the
lung pressure combined to the viscoelastic properties of the folds
and the Bernoulli effect \cite{CHAO5-1}. Normal voiced sounds appear
to be nearly periodic, although small perturbations are important
for the naturalness of  voice. In the last years  literature has
concerned with observation and characterization of nonlinear
phenomena in human and animal vocalization  \cite{toku,jasa108-4,facchini}.%
Simulations show that the two mass model reproduces behaviors
observed in natural vocalizations, including  some irregular
behaviors. Proposed by Ishizaka and Flanagan in 1972 \cite{IF}, it
was improved by Herzel and Titze in the last years
\cite{CHAO5-1,jasa94-6}. The model represents the vocal cords as a
viscoelastic coupled oscillator. There exists a substantial evidence
that vocal-fold vibration is a highly nonlinear process
\cite{Ndyn95-7,VFP143}, and the combined effects of nonlinear
biomechanical events and aerodynamic events can produce rich
irregular vibratory behaviors such as bifurcation and chaos
\cite{jasa110-4}. Furthermore, the two-mass model is able to produce
toroidal oscillations and chaotic transitions such as period
doubling and Ruelle-Takens-Newhouse. These behaviors are strongly
dependent on the variation of subglottal pressure and on the values
of the elastic constants \cite{jasa97-3}.
 In the paper \cite{AMR46-7}, Titze reported the equation of the
natural oscillation frequencies $\omega_i$ of the model:
\begin{equation}
\omega_{1,2}=\sqrt{\frac{k_{1}+k_{c}}{m_{1}}+\frac{k_{2}+k_{c}}{m_{2}}\pm
\sqrt{\left(\frac{k_{1}}{m_{1}+m_{2}}\right)^{2}+\left(\frac{k_{2}}{m_{1}+m_{2}}\right)^{2}
} }
\end{equation}
Using the standard parameters suggested by Ishizaka and Flanagan he
founds $f_{1}=120$ Hz and $f_{2}=220$ Hz.  These values are
inversely proportional to the weight of the vocal folds $m_i$ and
proportional to the elastic constants $k_i$.

According to Barr et al. \cite{barr}, the crying of the newborns is
simultaneously a sign, a symptom and a signal. It is the infant's
earliest form of communication but the significance of neonatal
crying is still unclear. Pain scales \cite{carbajal97} have recently
been developed to discriminate levels of pain suffered by newborns
but the direct relationship between  the nature of crying  and
effective pain suffered is rarely considered in medical studies.

Chandre \cite{PHYD181} showed that it is possible to describe and
identify chaotic and irregular dynamics from spectrograms of
irregular signals, while Mende \cite{Mende90} observed several
irregular behaviors in spectrograms of newborns cries. Subharmonics
and chaotic patterns have been found in cries of healthy infants as
well as infants with various perinatal complications \cite{PORL49}.
However the frequency and duration of episodes depended on the
health status of the newborn. This may indicate that brain control
on the vocal apparatus is poor or not completely developed
\cite{Mende90}. Dysphonia in infants was commonly related to several
pathologies of the vocal cords, and a complete classification of
these phenomena was made by Hirschberg \cite{PORL49-A}. Recently,
Robb showed that phenomena like Biphonation, Harmonic Doubling, and
F0 shift are common in healthy newborns too \cite{FPL03-55}.%

In this study we performed  a time-frequency analysis on the
vocalizations in order to find a relation between pain suffered and
the presence of irregular and chaotic dynamics in the signal. We
assessed whether onset of potentially chaotic cry segments, present
in spectrograms as noise concentration patterns (NP) was also
related to the degree of pain suffered by newborns. A hypothesis is
that onset of nonlinear phenomena  may be correlated with the muscle
contraction typical of suffering. Contraction causes large
variations in the elastic constants of the tissues and determines
different oscillating behaviors of the vocal folds.
\section{Materials and Methods}
Babies enrolled in this study were all healthy term babies; the
painful procedure was a heel-prick performed between 24 and 48 hours
of live for clinical screening purposes (to assess phenylalanine and
thyroid hormones plasma levels), and executed by two nurses
following a standardized procedure.

Each baby was recorded with a Sony video camera, and a pain score
was assigned on the DAN scale (Douleur Aigu\"{e} du Nouveau N\'e), a
validated scale whose reliability (specificity, sensibility,
accuracy and clinical feasibility) has been
assessed\cite{carbajal97}. The DAN score is obtained by scoring
three items  according to the extent they are present during the 30
seconds following the painful event: crying, arm and leg movements,
grimacing. The pain scale is explained   in table \ref{DAN} and the
total score ranges from 0 (minimum pain) to 10 (maximum pain).
%
%\begin{verbatim}
%Rispondere alle obiezioni del referee per quello che riguarda %
%l'affidabilita' della scala DAN. Punto 5 della lettera. Inoltre%
%spiegare quale parte del pianto e' stata presa in considerazione e
%perche'. Forse e' anche da spiegare il metodo con cui il sangue
%viene prelevato.
%\end{verbatim}
Recordings  from 40 newborns were divided into three groups,
depending on their assigned DAN score:
\begin{enumerate}
\item Low pain, DAN=1-4. A group of 10 newborns.
\item Medium pain, DAN=5-7. A group of 15 newborns.
\item High pain, DAN=8-10. A group of 15 newborns.
\end{enumerate}
The acoustic signals were obtained by an analog recorder and time
series of about 900 msec., corresponding to the first cry and to the
exact moment of the cut, were extracted from each recording. Both
power spectrum (PS) and phonograms were computed using a window of
1024 samples and a superposition region of 512 samples. For
completeness, for every cry episode, we have collected three bawls,
giving a greater importance to the first one, and using the other
two for completeness. In fact, only the first cry is directly
related to the painful event. We did not considered the successive
cries because they  were far from the sample, the effect of  pain
was reduced, and the successive shouts might be related to fear or
angry.

\section{Results}
All the spectra were characterized by high values of the harmonics,
with frequencies sometimes over the 10KHz, uncommon in adults, but
explainable considering that the natural oscillation frequencies are
inversely proportional to the masses of the vocal folds. We were
interested in the detection of the irregular patterns known in
literature as {\em Noise concentration patterns} (NP), directly
connected to the onset of chaotic oscillations \cite{HerzNAPD}.
Other behaviors, such as vibrato, biphonation and $F0$ shift, that
sometimes arise in newborns and infants cry and non-cry vocalization
were not considered as irregular in this particular context.
\subsection{Low DAN (1-4)}
\begin{figure}
\begin{center}
\includegraphics[height=6cm, keepaspectratio=true]{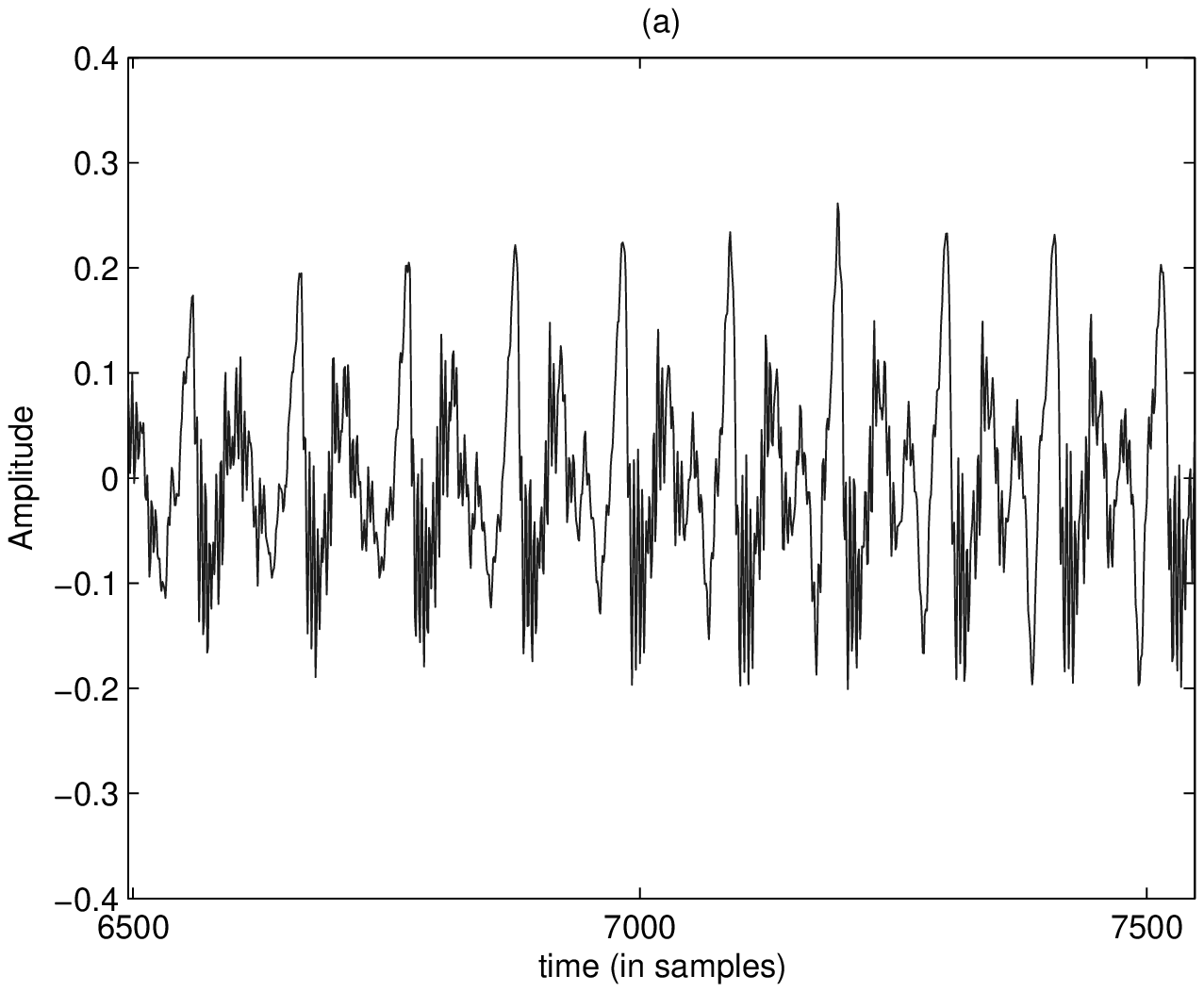}
\includegraphics[height=6cm, keepaspectratio=true]{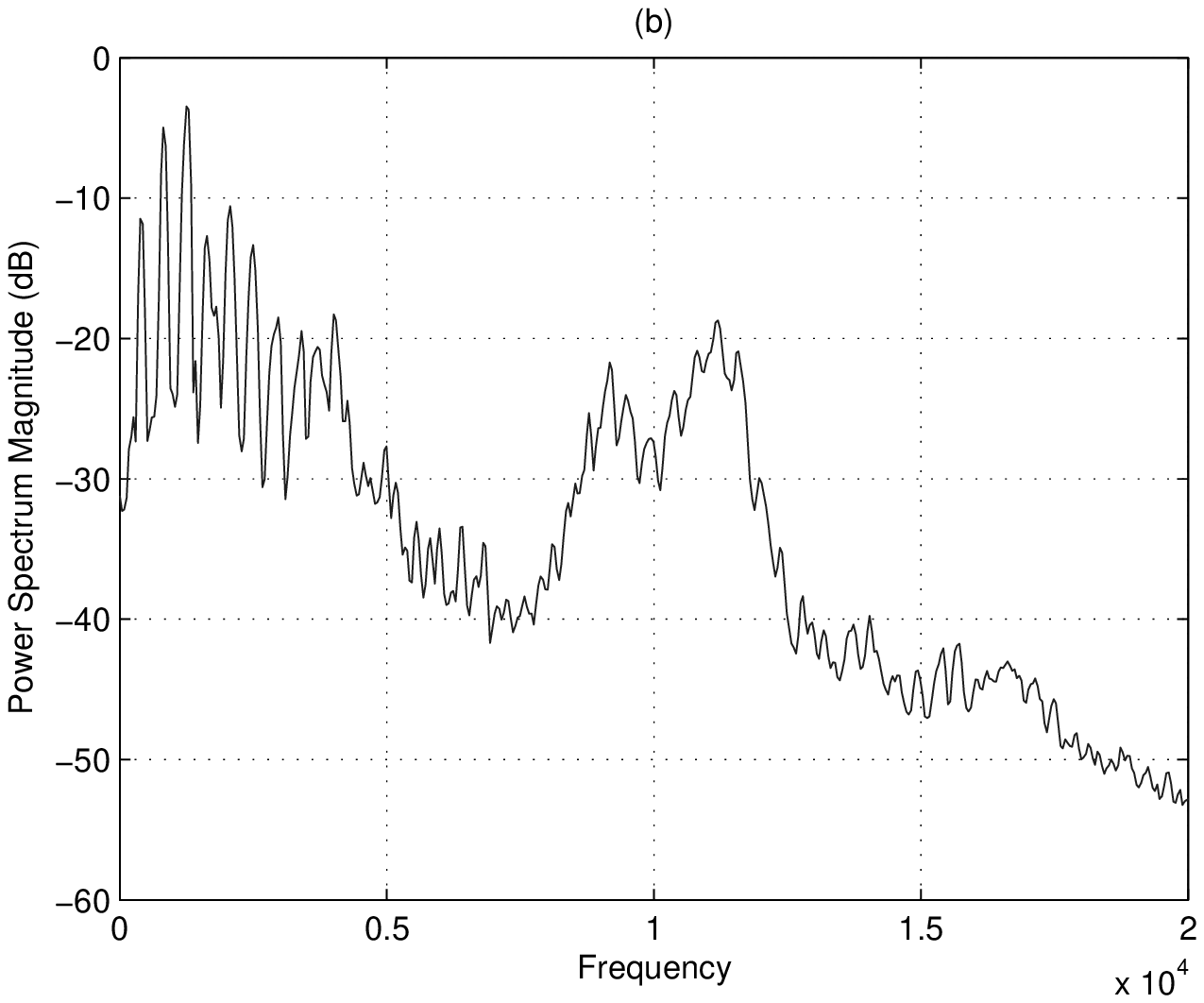}
\includegraphics[height=6cm, keepaspectratio=true]{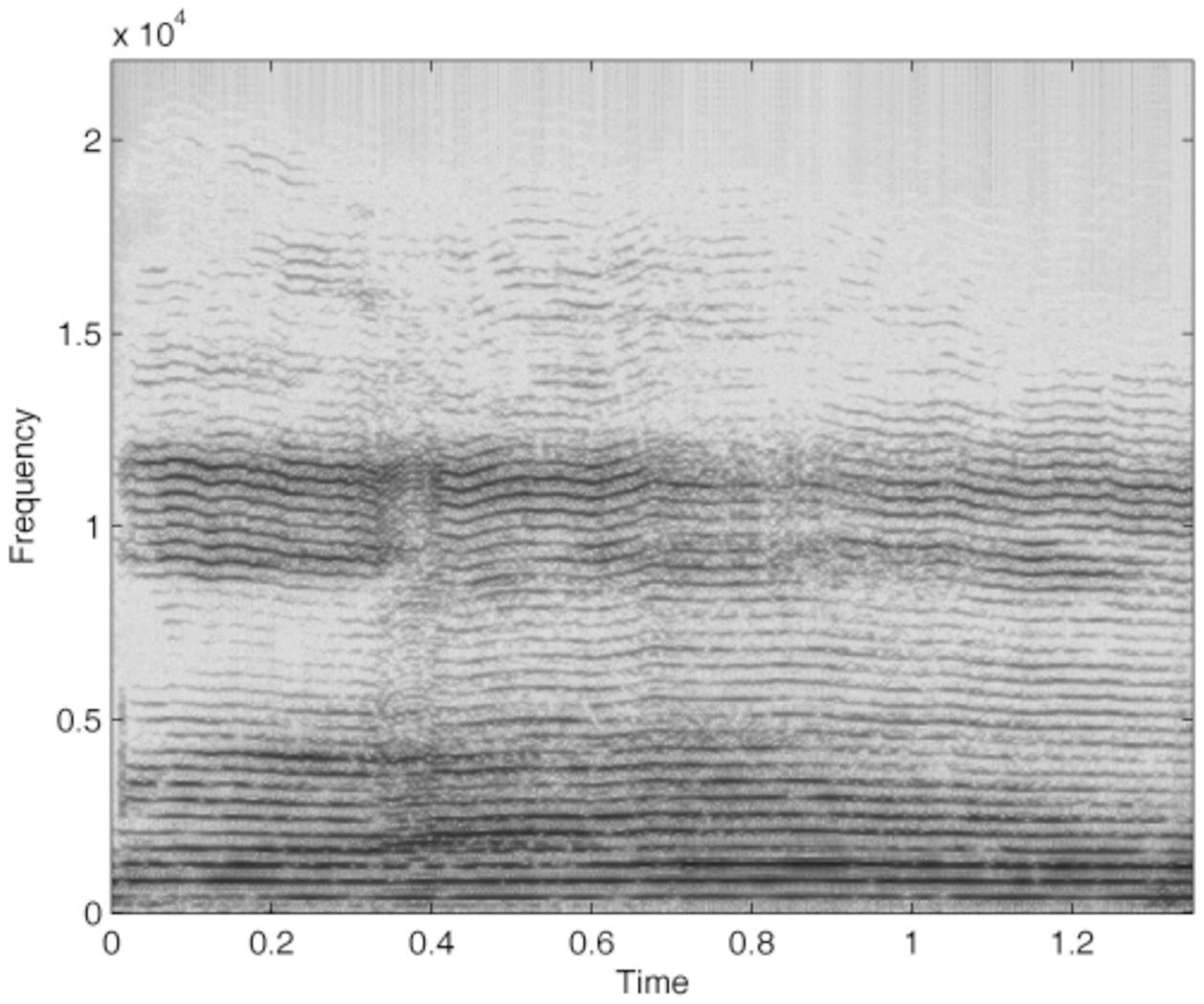}
\end{center}
% Here is how to import EPS art
\caption{Time and frequency analysis of a low DAN assessed cry.%
(a)A portion of about 1000 samples (22 msec.) showing the complex,
but periodic, nature of the signal.%
(b),(c) Both PS and spectrogram confirm the periodic nature of the
cry, the fundamental frequency is about 360 Hz, while some harmonics
are over the 10 KHz.} {\label{fig:DB}}
\end{figure}

Figure \ref{fig:DB} shows the time series, the power spectrum and
the spectrogram of a cry assessed with a low DAN score. The figure
is representative of the other cries analyzed. The power spectrum
and the spectrogram reveal the periodic nature of the vocalization
emitted. In particular, phenomena such BP and $F0$ shift are present
in all the other signals, but noise patterns are completely missing.
Summarizing we have inspected 10 cries from 10 different newborns:
all the cries were periodic, and no NP were detected. The inspection
of the successive shouts showed the same regular patterns.
\subsection{Medium DAN (5-7)}
\begin{figure}
\begin{center}
\includegraphics[height=6cm, keepaspectratio=true]{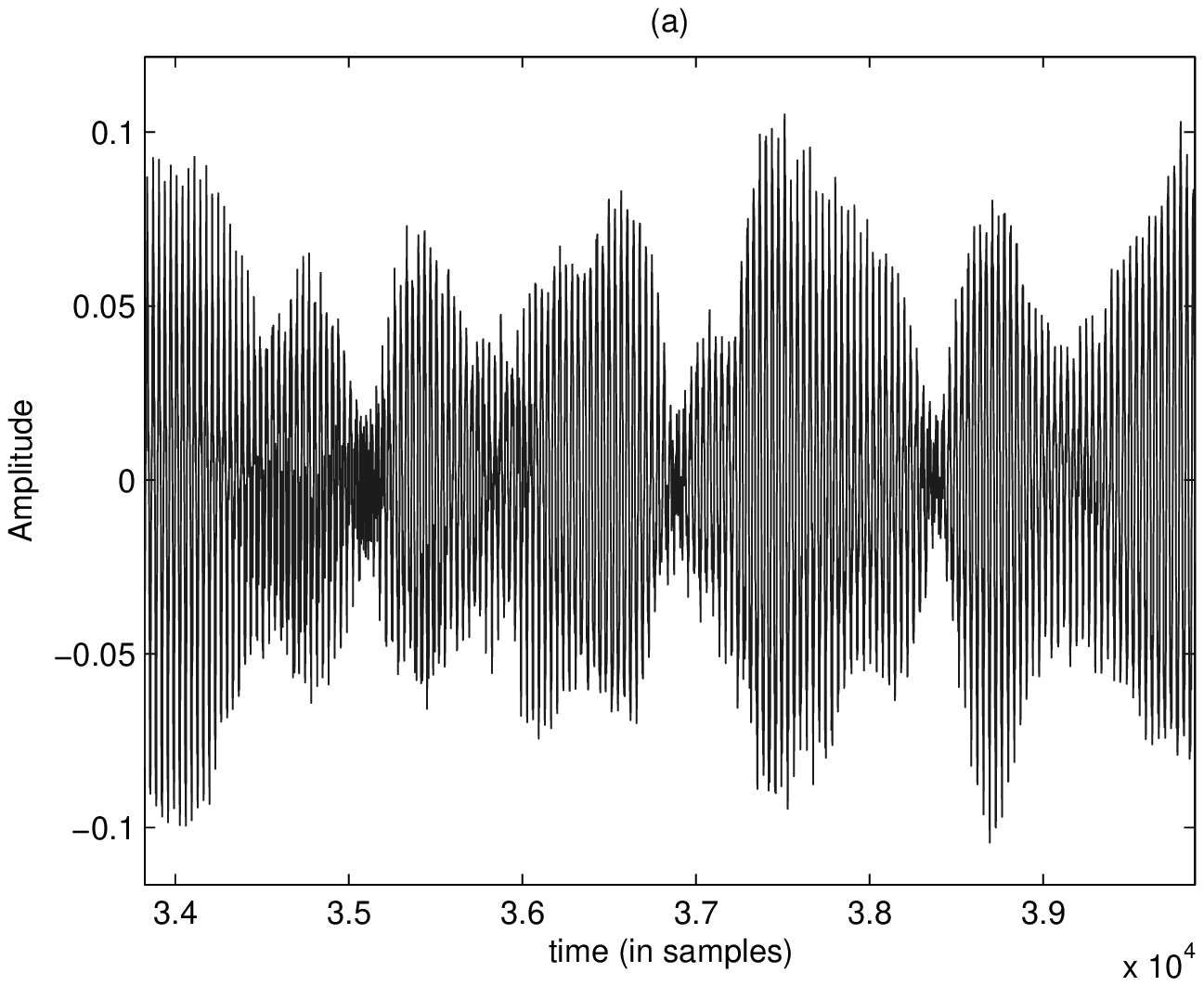}
\includegraphics[height=6cm, keepaspectratio=true]{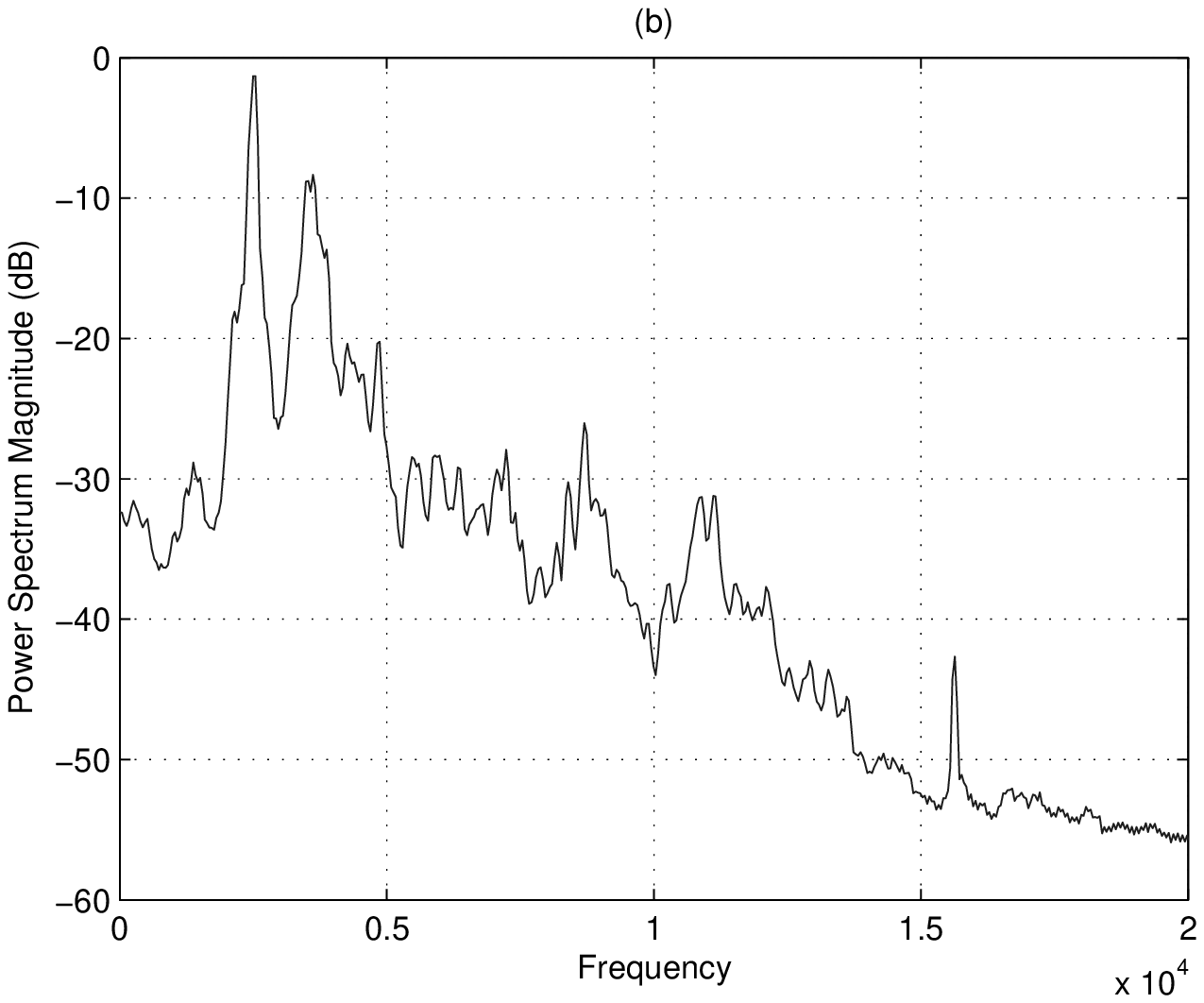}
\includegraphics[height=6cm, keepaspectratio=true]{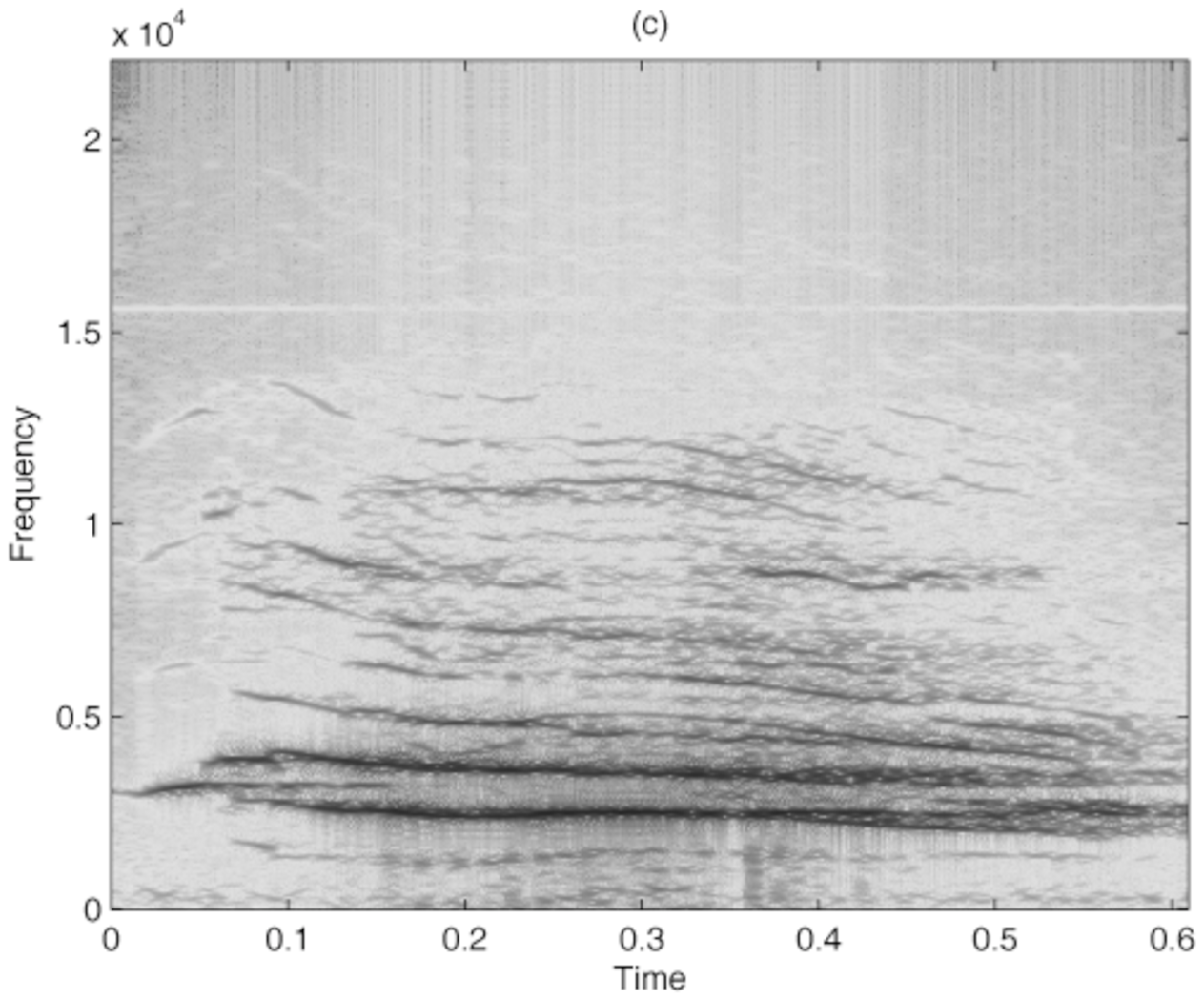}
\end{center}
% Here is how to import EPS art
\caption{Time and frequency analysis of a medium DAN assessed cry.%
(a) The signal looks more irregular, and the amplitude modulation is
a symptom of a torus-2 oscillation. (b) In the PS are visible two
main frequencies: $f_1=245$ Hz and $f_2=1403$ Hz and their linear
combinations $2f_2 - f_1$ and $2f_2 +3f_1$. This is usually referred
to toroidal oscillations. (c) The spectrogram shows the two main
frequencies and sob light concentrations of noise patterns. }
 {\label{fig:DM}}
\end{figure}

The main feature of this class of cries is the drastically deduced
number of harmonics and, as shown in figure \ref{fig:DM} the
presence of two different main frequencies, $f_1= 245$ Hz and
$f_2=1403$ Hz, their linear combinations $2f_2 - f_1$ and $2f_2
+3f_1$, and some intermittent concentrations of NP. In time, as
shown in figure \ref{fig:DM}(a), the pattern of the signal is more
aperiodic, and the amplitude variations are an effect of the
irregular oscillations. The presence of two different frequencies is
usually related to aperiodic, but not chaotic, oscillations. This
behavior is known as {\em torus-2} oscillation, and is the first
step of the Ruelle-Takens-Newhouse transition \cite{RTN87}, also
reported for the vocal cords \cite{HerzNAPD}. All the cries
belonging to this class have the same characteristics discussed
above.
\subsection{High DAN (8-10)}
\begin{figure}
\begin{center}
\includegraphics[height=6cm, keepaspectratio=true]{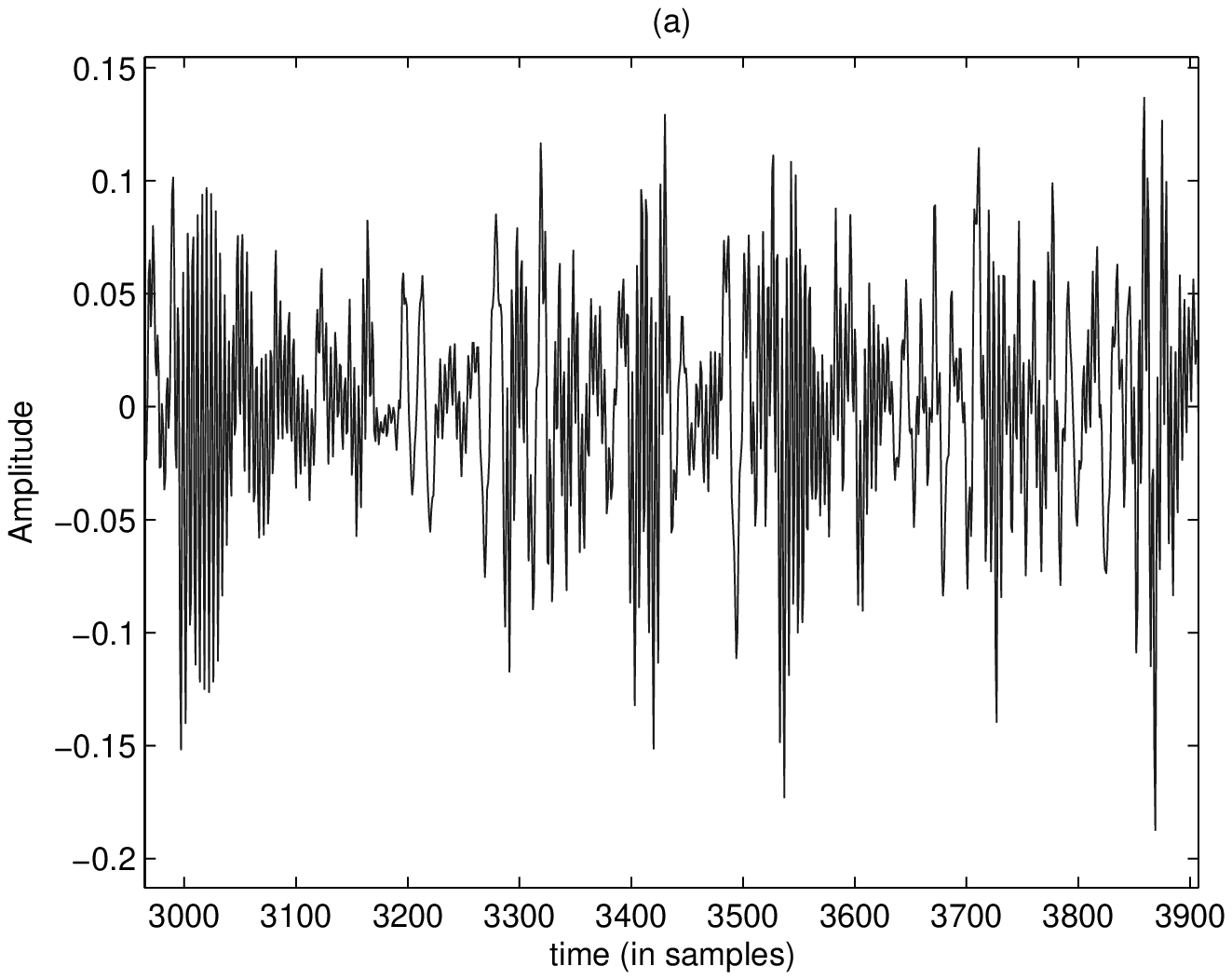}
\includegraphics[height=6cm, keepaspectratio=true]{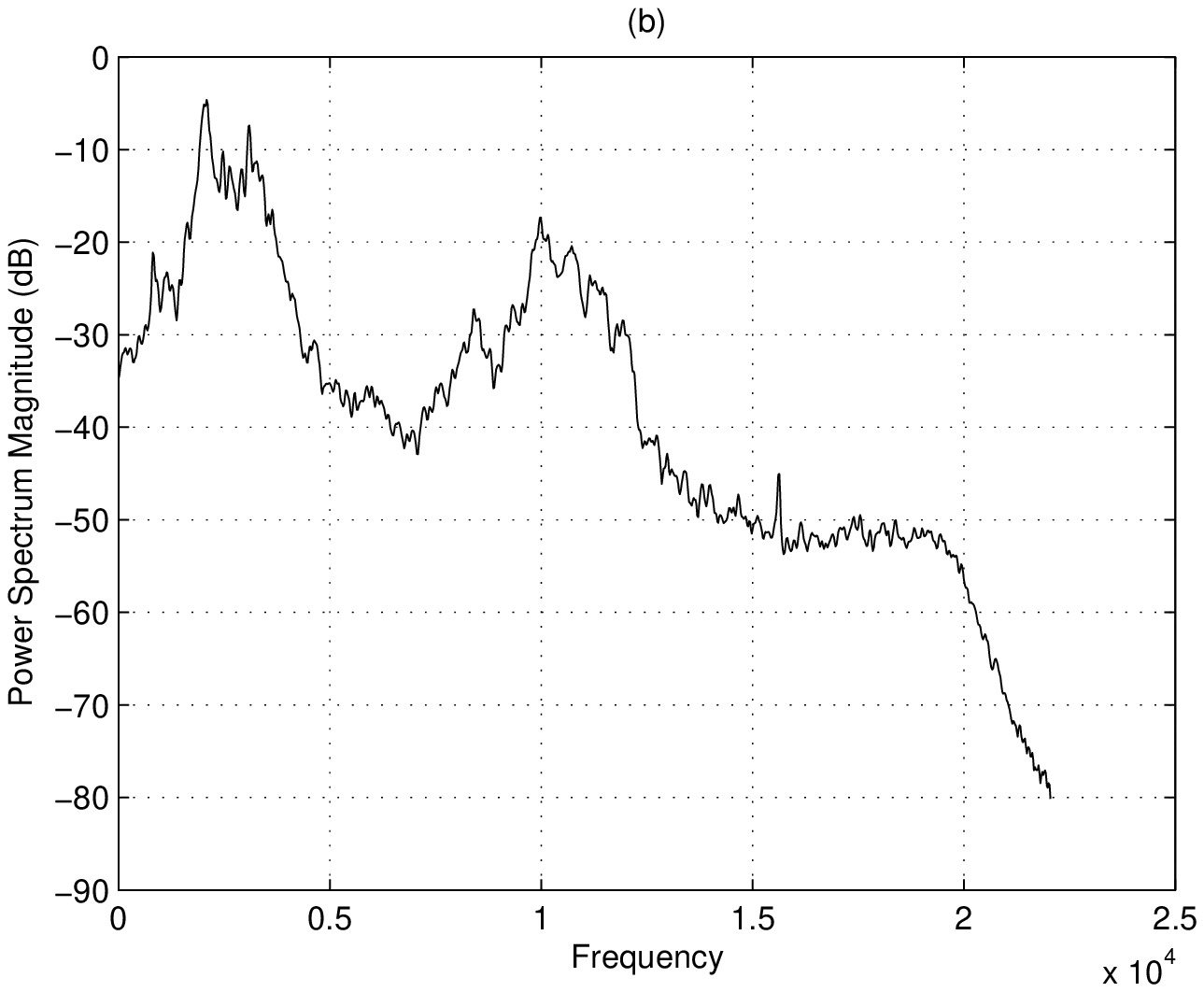}
\includegraphics[height=6cm, keepaspectratio=true]{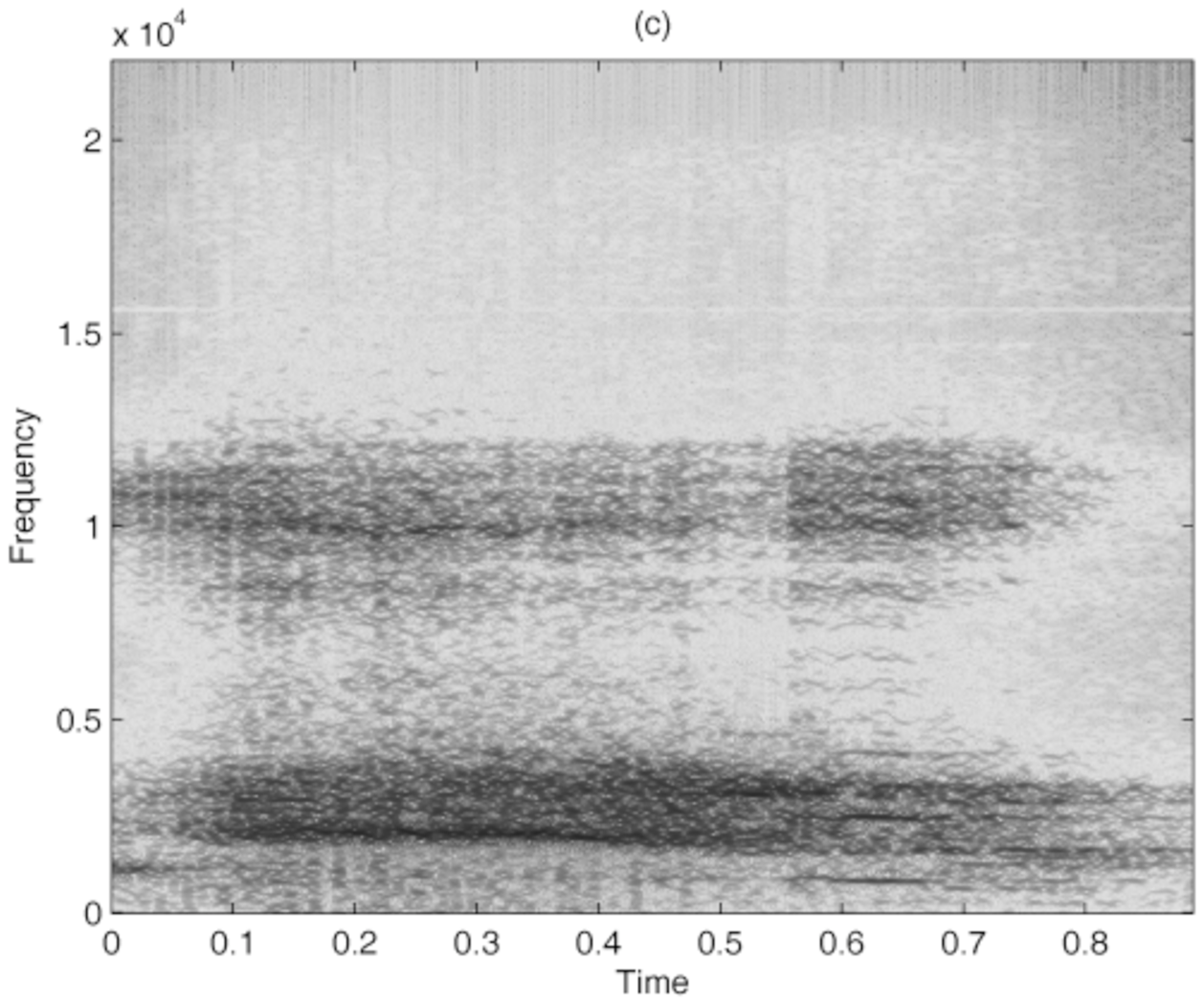}
\end{center}
% Here is how to import EPS art
\caption{Time and frequency analysis of an high DAN assessed cry.%
(a) The signal looks in time completely irregular, with
concentrations of high and low amplitude spikes and strong
variations in the period. (b),(c) Both PS and spectrogram confirm
the irregular nature of the signal, which may be considered chaotic.
In the PS is not possible to individuate any main frequency while
the two large peaks are referred to the two NP bands in the
spectrogram.} {\label{fig:DA}}
\end{figure}

When the DAN score is high, the spectra and the phonogram are like
those in Figure \ref{fig:DA}, showing large zones of NP without a
clear fundamental frequency. The appearance of the signal in time
(Figure \ref{fig:DA}(a)) is typical of irregular oscillations, and
both in PS and spectrogram (Figures \ref{fig:DA} (b) and (c)) are
visible large concentrations of NP around two large indistinct
frequency bands, usually referred to chaotic segments. All the cries
belonging to this class have the same spectrographic
characteristics.

\begin{figure}
\begin{center}
\includegraphics[height=8cm, keepaspectratio=true]{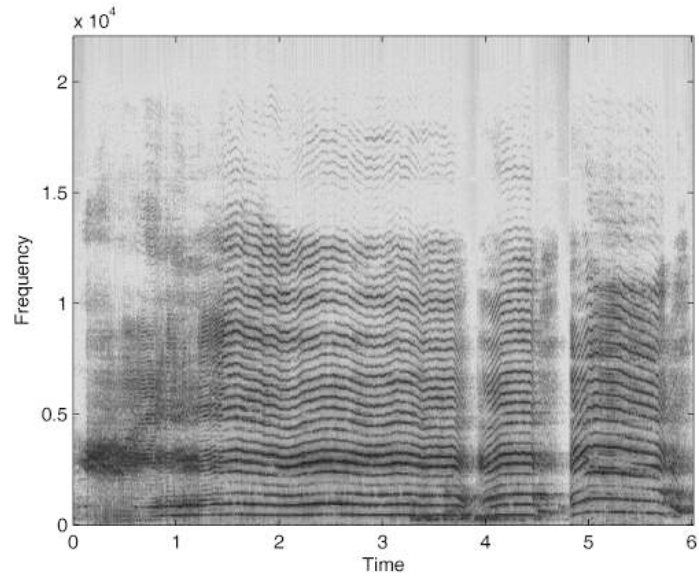}
\end{center}
% Here is how to import EPS art
\caption{Observation of the inverse transition
\emph{chaos$\to$torus-2$\to$limit cycle} in the spectrogram of a 6
seconds high DAN cry. Here are reported the first three shouts while
the moment of the sampling cut is at 0 seconds. In the time window
0-900 ms are only visible the two NP bands. In the region 900-1400
msec. are clearly visible the two main frequencies of the toroidal
oscillations, after that, the cry becomes regular. This inverse
transition is due to the effect of cry on the organism.}
{\label{fig:DA-cont}}
\end{figure}

Summarizing, the increase of pain induces a transition in the
oscillation of the vocal folds of the type limit {\em limit
cicle$\to$torus 2$\to$chaos}. Since the onward transition (caused by
pain increase) is not obtainable for the same newborn, we performed
a spectrogram analysis on the high DAN assessed cries, in order to
find the inverse transition. Figure \ref{fig:DA-cont} reports the
spectrogram of a 6 seconds cry consisting of three shouts. The
recording begins with the sampling cut (0 sec.). The first shout is
4 seconds long, while the other two are about 1 second long. It must
be pointed out that in the first shout is clearly visible the
inverse transition from NP (0-900 msec.) to periodic behavior (from
about 1.4 sec.) through a torus-2 oscillation (900 ms-1.4 sec.).
\section{Conclusions}
It was possible to discern pain levels from the spectrograms. For
DAN scores under 4 the vocalizations were characterized by periodic
behavior. At medium DAN  scores (DAN=5-7) the oscillation became
more irregular, and the presence of toroidal oscillation was
observed. At high DAN (DAN=8-10) values spectrograms were
characterized by large bands of Noise concentration Patterns.

If we consider variations in the bifurcation parameters of the vocal
folds system, especially lung pressure and the elastic constants of
vocal fold tissues, and assume that pain causes muscle contraction
and an increase in lung pressure \cite{bellieni}, we can explain
these results as distortion of the oscillating system caused by
large variations in vocal fold  tension and tissues constants of
distressed babies. According to this hypothesis, an increase in pain
induces a chaotic transition in vocal fold oscillation of the
newborn, so that crying associated with pain is strictly related to
noise patterns in the spectrograms. The inverse transition, induced
by the decrease of pain, was observed in the cries of highly
suffering newborns. Moreover, we proposed a method for the
assessment of pain based only on the spectral characteristic of cry.
\section{Acknowledgments}
The authors are grateful to the referees for their important
suggestions and comments.

%GATHER{bam.bib}   % For Gather Purpose Only
\bibliography{BamarXiv}% Produces the bibliography via BibTeX.

\begin{thebibliography}{10}

\bibitem{barr}
R.G. Barr, B.~Hopkins, and J.A. Green.
\newblock {\em Crying as a sign, a symptom, and a signal.}
\newblock Cambridge Univ. Press, 2000.

\bibitem{bellieni}
C.V. Bellieni, R.~Sisto, D.M. Cordelli, and G.~Buonocore.
\newblock Cry features reflect pain intensity in term newborns: an alarm
  threshold.
\newblock {\em Pediatric Research}, 55(1):142--146, January 2004.

\bibitem{carbajal97}
R.~Carbajal, A.~Paupe, E.~Hoenn, R.~Lenclen, and M.O. Martin.
\newblock Dan 1997 une \'{e}chelle comportamentale d'\'{e}valuation de la
  doleur aigue du nouveau-n\'{e}.
\newblock {\em Arch. Pediatr.}, 4:623--628, 1997.

\bibitem{PHYD181}
C.~Chandre, S.~Wiggins, and T.~Uzer.
\newblock Time-frequency analysis of chaotic systems.
\newblock {\em Physica D}, 181(3-4):171--196, August 2003.

\bibitem{facchini}
A.~Facchini, S.~Bastianoni, N.~Marchettini, and M.~Rustici.
\newblock Characterization of chaotic dynamics in the vocalization of the {\em
  cervus elaphus corsicanus}.
\newblock {\em J. Acoust. Soc. Am.}, 114(6):3040--3043, December 2003.

\bibitem{IF}
K.~Ishizaka~J.L. Flanagan.
\newblock Synthesis of voiced sounds from a two-mass model of the vocal cords.
\newblock {\em Bell Syst. Tech. J.}, 51:1233--1268, 1972.

\bibitem{AMR46-7}
H.~Herzel.
\newblock Bifurcations and chaos in voice signals.
\newblock {\em Appl. Mech. Rev.}, 46(7):399--413, July 1993.

\bibitem{CHAO5-1}
H.~Herzel, D.~Berry, I.~Titze, and I.~Steincke.
\newblock Nonlinear dynamics of the voice: Signal analysis and biomechanical
  modeling.
\newblock {\em CHAOS}, 5(1):30--34, 1995.

\bibitem{HerzNAPD}
H.~Herzel, J.~Holzfuss, Z.J. Kowalik, B.~Pompe, and R.~Reuter.
\newblock Detecting bifurcations in voice signals.
\newblock In H.~Kantz, J.~Kurths, and G.~Mayer-Kress, editors, {\em Nonlinear
  analysis of physiological data}, pages 325--344. Springer, 1998.

\bibitem{Ndyn95-7}
H.~Herzel and C.~Knudsen.
\newblock Bifurcations in a vocal fold model.
\newblock {\em Nonlinear Dyn.}, 7:53--64, 1995.

\bibitem{PORL49-A}
J.~Hirschberg.
\newblock Dysphonia in infants.
\newblock {\em Int. J. of Pediatric Othorinolaryngology}, 49S(4):S293--S296,
  1999.

\bibitem{jasa110-4}
J.J. Jiang, Y.~Zhang, and J.~Stern.
\newblock Modeling of chaotic vibrations in symmetric vocal folds.
\newblock {\em J. Acoust. Soc. Am.}, 110(4):2120--2128, October 2001.

\bibitem{jasa94-6}
J.C. Lucero.
\newblock Dynamics of the two-mass model of the vocal folds: Equilibria,
  bifurcations and oscillation region.
\newblock {\em J. Acoust. Soc. Am.}, 94(6):3104--3111, December 1993.

\bibitem{Mende90}
W.~Mende, H.~Herzel, and K.~Wermke.
\newblock Bifurcations and chaos in newborn infant cries.
\newblock {\em Physics lett. A}, 145(8,9):418--424, April 1990.

\bibitem{PORL49}
K.~Michelsson and O.~Michelsson.
\newblock Phonation in the newborn, infant cry.
\newblock {\em Int. J. of Pediatric Othorinolaryngology}, 49S(1):S297--S301,
  199.

\bibitem{RTN87}
S.~Newhouse, D.~Ruelle, and F.~Takens.
\newblock Occurrence of strange axiom a attractors near quasiperiodic flows on
  $ t^m , m\ge 3$.
\newblock {\em Commun. Math. Phys.}, 64:35--40, 1987.

\bibitem{jasa108-4}
T.~Riede, H.~Herzel, D.~Mehwald, W.~Seidner, E.~Trumler, G.~B\"{o}hme, and
  G.~Tempbrock.
\newblock Nonlinear phenomena in the natural howling of a dog-wolf mix.
\newblock {\em J. Acoust. Soc. Am.}, 108(4):1435--1442, October 2000.

\bibitem{FPL03-55}
M.P. Robb.
\newblock Bifurcations and chaos in the cries of full-tem and preterm infants.
\newblock {\em Folia Phoniatrica et Logopaedica}, 55:233--240, 2003.

\bibitem{jasa97-3}
I.~Steincke and H.~Herzel.
\newblock Bifurcations in asymmetric vocal fold model.
\newblock {\em J. Acoust. Soc. Am.}, 97(3):1874--1884, March 1995.

\bibitem{VFP143}
I.R. Titze, R.~Baken, and H.~Herzel.
\newblock Evidence of chaos in vocal fold vibration.
\newblock In I.R. Titze, editor, {\em Vocal Fold Physiology: New Frontier in
  Basic Science}, pages 143--188, 1993.

\bibitem{toku}
I.~Tokuda, T.~Reide, J.~Neubauer, M.J. Owren, and H.~Herzel.
\newblock Nonlinear analysis of irregular animal vocalizations.
\newblock {\em J. Acoust. Soc. Am.}, 111(6):2908--2919, june 2002.

\end{thebibliography}
\bibliographystyle{plain}

\newpage

\renewcommand{\baselinestretch}{1}
\begin{table}
  \centering
  \caption{DAN: behavioral acute pain-rating scale for neonates}\label{DAN}
  \begin{tabular}{ll}
    % after \\: \hline or \cline{col1-col2} \cline{col3-col4} ...
    \hline
    Measure & Score \\
    \hline
    {\bf Facial Expressions} &  \\
    Calm & 0\\
    Snivels and alternates gentle eye opening and closing & 1 \\
    Determine intensity of one or more of eye squeeze, &  \\
    brow bugle nasolabial furrow &  \\
    Mild, intermittent with return to calm  & 2 \\
    Moderate & 3 \\
    Very pronounced, continuous & 4 \\
    \hline
    {\bf Limb movements} &  \\
    Calm or gentle movements & 0 \\
    Determine intensity of one or more of &  \\
    the following signs: pedals, toes spread legs tensed and pulled up, &  \\
     agitation of arms, withdrawal reaction&  \\
    Mild intermittent with return to calm & 1 \\
    Moderate & 2 \\
    Very pronounced, continuous & 3 \\
    \hline
    {\bf Vocal expressions} &  \\
    No complaints & 0 \\
    Moans briefly; for intubated child, looks anxious or uneasy  & 1 \\
    Intermittent crying; for intubated child,& 2 \\
     gesticulations of intermittent crying & \\
    Long-lasting crying, continuous howl; & 3 \\
    for intubated child gesticulation of continuous crying & \\
    \hline
  \end{tabular}
\end{table}

\renewcommand{\baselinestretch}{1.5}
\newpage

%%%% FIGURE

\newpage
% figura del pianto a DAN BASSO

\newpage

%figura del Pianto a DAN MEDIO

% figura del pianto a DAN ALTO

%^ secondi di pianto a DAN ALTO

\end{document}